\documentclass{elsart}

\usepackage{graphicx}

\usepackage{amssymb}

\begin{document}
\begin{frontmatter}
\title{Measurement-induced disturbance and thermal negativity of qutrit-qubit mixed spin chain}
\author[label1]{Guo-Feng Zhang\corauthref{cor1}},
\corauth[cor1]{Corresponding author:gf1978zhang@buaa.edu.cn}
\author[label2]{Yu-Chen Hou},
\author[label3]{Ai-Ling Ji}
\address[label1]{Department of Physics, School of Physics and Nuclear Energy Engineering, Beihang University, Xueyuan Road No. 37, Beijing 100191, PR China}
\address[label2]{School of Electronic and Information Engineering, Beihang University, Xueyuan Road No. 37, Beijing 100191, PR China}
\address[label3]{State Key Laboratory for Surface Physics, Institute of Physics, Chinese Academy of Sciences
P. O. Box 603, Beijing, 100190, PR China}

\begin{abstract}
\quad We investigate the quantum correlation in a qutrit-qubit mixed spin chain based on measurement-induced disturbance (MID) [S. Luo, Phys. Rev. A, 77, (2008) 022301]. We also compare MID and thermal entanglement measured by negativity and illustrate their different characteristics.
\end{abstract}

\begin{keyword}
D. Measurement-induced disturbance (MID); D. Negativity; A. Qutrit-qubit mixed spin chain
\end{keyword}
\end{frontmatter}

\section{Introduction}

Quantum entanglement and quantum correlation are two fascinating quantities in quantum world. They play a central role in revealing the feature of quantum physics. Quantum entanglement, which is important to quantum information processing, has been studied widely while quantum correlation seems to have been seldom exploited before, especially for the solid spin system. Quantum correlation arises from noncommutativity of operators representing states, observables, and measurements \cite{sll}. Quantum entanglement can be realized in many kinds of physical systems which involve quantum correlation. Now quite a few people take it for granted that quantum entanglement is quantum correlation. Neverless, Li and Luo illustrated through simple examples that the entanglement of formation may exceed the total correlations as quantified by the quantum mutual information \cite{nli}. Several authors have pointed out that in some quantum tasks which cannot be simulated by classical methods, it is the correlations (of course, of a quantum nature), rather than entanglement, that are responsible for the improvements. Now we recognize that quantum entanglement is a special kind of quantum correlation, but not the same with quantum correlation. So, it is very interesting and necessary to study the relation between quantum entanglement and quantum correlation.

Solid spin systems are the natural candidates for the realization of the entanglement compared with the other physical systems. Recently, there is a growing interest in the study of thermal entanglement. This is motivated by the fact that qubits composed of any physical system are carried experimentally out at finite temperatures, but thermal effects will cause disentanglement of entangled qubits. It is necessary to take the effects of temperature into quantum information account. A lot of interesting work about thermal entanglement in spin systems have been done \cite{man,xwa,glk,kmo,gfz1,yye,dvk,gfz2,xgw,shi,mca}. The previous studies about spin chain entanglement is mostly for Heisenberg spin $1/2$ system. The spin of most of system is greater than $1/2$, such as inorganic compounds $ACu(pbaOH)(H_{2}O_{3})\cdot2H_{2}O$ \cite{kon}, each of which unit cells contains two different mixed-spin (S, $1/2$). Therefore, to study mixed spin chain has an important significance.

The classification of correlations based on quantum measurements has arisen in recent years \cite{mpi,slu,nli2}. In particular, quantum discord (QD) as a measure of quantum correlations, initially introduced by Ollivier and Zurek \cite{hol} and by Henderson and Vedral \cite{lhe} is attracting increasing
interest \cite{jcu,mdl,jsx,gad,rva,pgi,rcg,dos,mzw}. Recently, some authors \cite{twe} have pointed out that thermal quantum discord (TQD), in contrast to entanglement and other thermodynamic quantities, spotlight the critical points associated with quantum phase transitions (QPTs) for some spin chain model even at a finite temperature $T$. They think that the remarkable property of TQD is an important tool that can be readily applied to the reduction of the experimental demands to determine critical points for QPTs.

Unlike QD, Luo \cite{sll} introduced a classical vs quantum dichotomy in order to classify and quantify statistical correlations in bipartite states. In this paper, we will explore quantum correlation based on Luo's method \cite{sll} and investigate the dependences of spin-spin coupling and external magnetic field on quantum correlation in a qutrit-qubit system. The comparison between quantum correlation and thermal entanglement measured by negativity will be given.

Our paper is arranged as follows: in Sec.2, the model and the solution will be given. We will calculate MID and thermal negativity in Sec.3 and give a comparison. In Sec.4, we give a summary for our main results.

\section{The model solution and thermal state}
We consider a qutrit-qubit mixed spin chain $(1,1/2)$ in an applied magnetic field of the form
\begin{equation}
\emph{H}=JS_{1}\cdot s_{2}+Bs^{z}_{2},
\end{equation}
where $J$ is the real coupling coefficients between spin-$1$ and spin-$1/2$ particles, $B$ is magnetic field which is only applied to $z$-direction of spin-$1/2$ particle. $S_{1}$ and $s_{2}$ are the spin operators associated with the two particles.  $S_{1}^{\alpha}$$(\alpha=x,y,z)$ denotes spin-$1$ operator, its components take the form,
\begin{equation}
S_{1}^{x}=\frac 1{\sqrt{2}}\left(
\begin{array}{lll}
0 & 1 & 0 \\
1 & 0 & 1 \\
0 & 1 & 0
\end{array}
\right),\nonumber
\\S_{1}^{y}=\frac 1{\sqrt{2}}\left(
\begin{array}{lll}
0 & -i & 0 \\
i & 0 & -i \\
0 & i & 0
\end{array}
\right), \nonumber
\\S_{1}^{z}=\left(
\begin{array}{lll}
1 & 0 & 0 \\
0 & 0 & 0 \\
0 & 0 & -1
\end{array}
\right),
\end{equation}
$s_{2}^{\alpha}$$(\alpha=x,y,z)$ is Pauli matrix and $B\geq0$ is restricted. Note that we are working in units so that $J$ and $B$ are dimensionless.

To evaluate MID and thermal negativity we first of all find the
eigenvalues and the corresponding eigenstates of the Hamiltonian (1) which are seen to be
\begin{eqnarray}
H|\psi_{1}\rangle&=&(J-B)|\psi_{1}\rangle,
H|\psi_{2}\rangle=(J+B)|\psi_{2}\rangle,\nonumber
\\H|\psi_{3}^{\pm}\rangle&=&-\frac{a_{\pm}}{2}|\psi_{3}^{\pm}\rangle,
H|\psi_{4}^{\pm}\rangle=-\frac{b_{\pm}}{2}|\psi_{4}^{\pm}\rangle,
\end{eqnarray}
where $a_{\pm}=J\pm\sqrt{4B^{2}-4BJ+9J^{2}}$,
$b_{\pm}=J\pm\sqrt{4B^{2}+4BJ+9J^{2}}$.
And the corresponding eigenstates are explicitly given by
\begin{eqnarray}
|\psi_{1}\rangle&=&|-1,0\rangle, |\psi_{2}\rangle=|1,1\rangle,\nonumber
\\|\psi_{3}^{\pm}\rangle&=&\frac{1}{\sqrt{1+c_{\mp}^{2}}}(|-1,1\rangle+c_{\mp}|0,0\rangle),\nonumber
\\|\psi_{4}^{\pm}\rangle&=&\frac{1}{\sqrt{1+d_{\pm}^{2}}}(|0,1\rangle-d_{\pm}|1,0\rangle),
\end{eqnarray}
with $c_{\pm}=(a_{\pm}-2B)/(2\sqrt{2}J)$,
$d_{\pm}=(b_{\pm}+2B)/(2\sqrt{2}J)$. Here $\left|
x,y\right\rangle $ ($x=1,0,-1$ and $y=1,0$) are the eigenstates
of $S_{1}^{z}s_{2}^{z}$.

The state of a system at thermal equilibrium can be described by
the density operator $\rho (T)=\exp (-\beta H)/Z$, where
$Z=Tr[\exp (-\beta H)]$ is the partition function and $\beta
=1/k_BT$ ($k_B$ is Boltzmann's constant being set to be unit
$k_B=1$ hereafter for the sake of simplicity and $T$ is the
temperature). The entanglement in the thermal state is called
thermal entanglement. The density operator $\rho(T)$ can be expressed in
terms of the eigenstates and the corresponding eigenvalues as
\begin{equation}
\rho(T) =\frac 1Z\sum_{l}\exp [-\beta E_l]\left| \psi _l\right\rangle
\left\langle \psi _l\right|,
\end{equation}
where $E_l$ is the eigenvalue of the corresponding eigenstates
and the partition function is $Z=2e^{-J/T}[\cosh[B/T]+e^{(3J)/(2T)}(\cosh[\Lambda_{+}/(2T)]+\cosh[\Lambda_{-}/(2T)])]$ with $\Lambda_{\pm}=\sqrt{4B^{2}\pm4BJ+9J^{2}}$.

\section{The negativity and MID}

We will calculate the negativity and MID associated with the state (5) and give a detailed comparison between these two quantities.

\emph{Negativity. }The Peres-Horodecki
criterion \cite{ape} gives a qualitative way for judging if a
state is entangled. The quantitative version of the criterion was
developed by Vidal and Werner \cite{gvi}. They presented a measure
of entanglement called negativity that can be computed
efficiently, and the negativity does not increase under local
manipulations of the system. The negativity of a state $\rho $ is
defined as
\begin{equation}
N(\rho )=\sum_i\left| \mu _i\right|,
\end{equation}
where $\mu _i$ is the negative eigenvalue of $\rho ^{T_1}$, and
$T_1$ denotes the partial transpose with respect to the first
system. The negativity $N$ is related to the trace norm of $\rho
^{T_1}$ via\cite{gvi}
\begin{equation}
N(\rho )=\frac{\left| \left| \rho ^{T_1}\right| \right| _1-1}2.
\end{equation}
where the trace norm of $\rho ^{T_1}$ is equal to the sum of the
absolute values of the eigenvalues of $\rho ^{T_1}$. If $N>0$,
then the two-spin state is entangled.

For our purpose to evaluate the negativity in what following we
need to have a partially transposed density matrix $\rho ^{T_1}$
of original density matrix $\rho(T)$ with respect to the eigenbase
of any one spin particle ( say particle $1$) which is found in the
basis $\left| x,y\right\rangle $ ($x=1,0,-1$ and $y=1,0$) as
\begin{equation}
\rho ^{T_1}=\frac{1}{Z}\left(%
\begin{array}{ccccccccc}
  e^{(-B-J)/T} & 0 & 0 & \rho_{23} & 0 & 0  \\
  0 & \rho_{22} & 0 & 0 & 0 & 0 \\
  0 & 0 & \rho_{33} & 0 & 0 & \rho_{45}  \\
  \rho_{23} & 0 & 0 & \rho_{44} & 0 & 0  \\
  0 & 0 & 0 & 0 & \rho_{55} & 0  \\
  0 & 0 & \rho_{45} & 0 & 0 & e^{(B-J)/T}  \\
\end{array}%
\right),
\end{equation}
where
$\rho_{22}=d_{-}^{2}e^{b_{-}/(2T)}/(1+d_{-}^{2})+d_{+}^{2}e^{b_{+}/(2T)}/(1+d_{+}^{2})$,
$\rho_{33}=e^{b_{-}/(2T)}/(1+d_{-}^{2})+e^{b_{+}/(2T)}/(1+d_{+}^{2})$,
$\rho_{44}=c_{-}^{2}e^{a_{+}/(2T)}/(1+c_{-}^{2})+c_{+}^{2}e^{a_{-}/(2T)}/(1+c_{+}^{2})$,
$\rho_{55}=e^{a_{+}/(2T)}/(1+c_{-}^{2})+e^{a_{-}/(2T)}/(1+c_{+}^{2})$,
$\rho_{23}=-d_{-}e^{b_{-}/(2T)}/(1+d_{-}^{2})-d_{+}^{2}e^{b_{+}/(2T)}/(1+d_{+}^{2})$ and
$\rho_{45}=c_{-}e^{a_{+}/(2T)}/(1+c_{-}^{2})+c_{+}e^{a_{-}/(2T)}/(1+c_{+}^{2})$.

\emph{MID. }We can apply local measurement $\{\prod_{k}\}$($\prod_{k}\prod_{k^{'}}=\delta_{kk^{'}}\prod_{k}$ and $\sum_{k}\prod_{k}=1$) to any bipartite state $\rho$ (of course, including thermal state (5)). Here $\prod_{k}=\prod_{i}^{1}\otimes\prod_{j}^{2}$ and $\prod_{i}^{1}$, $\prod_{j}^{2}$ are complete projective measurements consisting of one-dimensional orthogonal projections for parties $1$ and $2$. After the measurement, we get the state $\prod(\rho)=\sum_{ij}(\prod_{i}^{1}\otimes\prod_{j}^{2})\rho(\prod_{i}^{1}\otimes\prod_{j}^{2})$ which is a classical state \cite{sll}. If the measurement
$\prod$ is induced by the spectral resolutions of the reduced states $\rho^{1}=\sum_{i}p_{i}^{1}\prod_{i}^{1}$ and $\rho^{2}=\sum_{j}p_{j}^{2}\prod_{j}^{2}$, the measurement leaves the marginal information invariant and is in a certain sense the least disturbing. In fact, $\prod(\rho)$ is a classical state that is closest to the original state $\rho$ since this kind of measurement can leave the reduced states invariant. One can use any reasonable distance between $\rho$ and $\prod(\rho)$ to measure the quantum correlation in $\rho$. In this paper, we will adopt Luo's method \cite{sll}, i.e., quantum mutual information difference between $\rho$ and $\prod(\rho)$,  to measure quantum correlation in $\rho$. The total correlation in a bipartite state $\rho$ can be well quantified by the quantum mutual information $I(\rho)=S(\rho^{1})+S(\rho^{2})-S(\rho)$, and $I(\prod(\rho))$ quantifies the classical correlations in $\rho$ since $\prod(\rho)$ is a classical state. Here $S(\rho)=$-tr$\rho$log$\rho$ denotes the von Neumann entropy, and the logarithm is always understood as base $2$ in this paper. So the quantum correlation can be quantified by the measurement-induced disturbance\cite{sll}
\begin{equation}
Q(\rho)=I(\rho)-I(\prod(\rho)).
\end{equation}

After simple calculations, we can get the reduced density matrix associated with the thermal state (5)
\begin{equation}
\rho^{1}=\frac{1}{Z}\left(
           \begin{array}{ccc}
             e^{(-B-J)/T}+\rho_{22} & 0 & 0 \\
             0 &\rho_{33}+\rho_{44} & 0 \\
             0 & 0 & \rho_{55}+e^{(B-J)/T} \\
           \end{array}
         \right),
\end{equation}
and
\begin{equation}
\rho^{2}=\frac{1}{Z}\left(
                       \begin{array}{cc}
                         e^{(-B-J)/T}+\rho_{33}+\rho_{55} & 0 \\
                         0 & e^{(B-J)/T}+\rho_{22}+\rho_{44} \\
                       \end{array}
                     \right).
\end{equation}

On the other hand, by taking $\prod_{i}^{1}=|i\rangle\langle i|(i=1,0,-1)$ and $\prod_{j}^{2}=|j\rangle\langle j|(j=1,0)$, we have
\begin{eqnarray}
\prod(\rho)=\frac{1}{Z}
\left(
  \begin{array}{cccccc}
    e^{(-B-J)/T} & 0 & 0 & 0 & 0 & 0 \\
    0 &  \rho_{22} & 0 & 0 & 0 & 0 \\
    0 & 0 &  \rho_{33} & 0 & 0 & 0 \\
    0 & 0 & 0 & \rho_{44} & 0 & 0 \\
    0 & 0 & 0 & 0 & \rho_{55} & 0 \\
    0 & 0 & 0 & 0 & 0 & e^{(B-J)/T} \\
  \end{array}
\right)
\end{eqnarray}
and
\begin{eqnarray}
[\prod(\rho)]^{1}=\rho^{1}, [\prod(\rho)]^{2}=\rho^{2},
\end{eqnarray}
From Eq.(7) and (8), thermal negativity can be obtained. Also, we can get MID from Eq. (5), (9) and (12).
\begin{figure}[h]
\begin{center}
\includegraphics[width=10 cm]{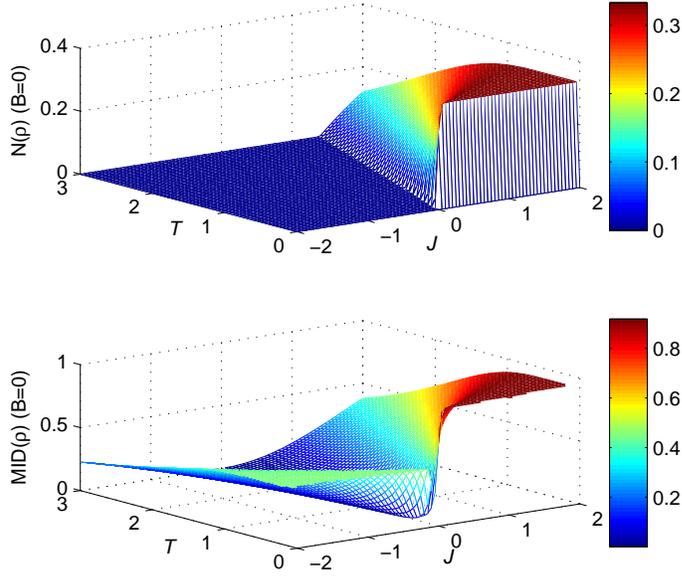}
\caption{(Color online) Negativity and MID for $B=0$ case. $T$ is plotted in units of the Boltzmann constant $k$. And we work in units where $B$ and $J$ are dimensionless.}
\end{center}
\end{figure}
\begin{figure}[h]
\begin{center}
\includegraphics[width=11 cm,height=6 cm]{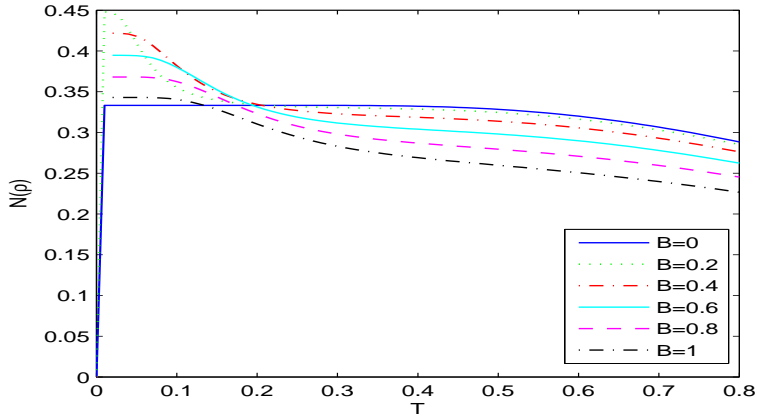}
\caption{(Color online) Quantum entanglement measured by negativity for any $B$ when $J=1$.}
\end{center}
\end{figure}

We perform the numerical simulation of the two quantities. The results for the external magnetic field $B=0$ are given in Fig.1. We find that negativity and MID evolve with respective to coupling constant and temperature very differently. There is no entanglement for $J<0$, i.e., ferromagnetic case, and a higher coupling constant will excite more entanglement for antiferromagnetic case. Moreover, the critical temperature above which entanglement is zero becomes higher for a strong coupling constant. Quantum correlation exists for both antiferromagnetic and ferromagnetic cases. The temperature always play a negative role in these two quantities, which can be easily understood since these two quantities are not classical. It is shown that negativity experiences a sudden transition when temperature changes from a finite value to zero, while quantum correlation evolves continuously with respective to temperature even it tends to be zero.

In order to see clearly the effects of magnetic field on these two quantities, we gave the evolution of negativity and MID in Fig.2 and fig.3 respectively for antiferromagnetic $J=1$ case. From Fig.2, we can see that for a certain magnetic field (for example $B=0.2$ in Fig.2), the system will have more entanglement at a temperature which is very near zero. There is a sudden transition for a weak magnetic field when the temperature changes from zero to a finite value while no sudden transition for a stronger magnetic field. We can analyze these results as follows. When $B$ is small, at zero temperature the system will be in an unentangled ground state. A lower temperature will excite entanglement. So, negativity will change from zero to a finite value. However, when $B$ is large, at zero temperature the system is entangled, thus negativity will evolve smoothly from a finite to another finite value. No sudden transition occurs for quantum correlation. MID has a minimal value at which the temperature is low but not zero for any magnetic field, and will tend to be zero when the temperature is too high.
\begin{figure}
\begin{center}
\includegraphics[width=11 cm,height=6 cm]{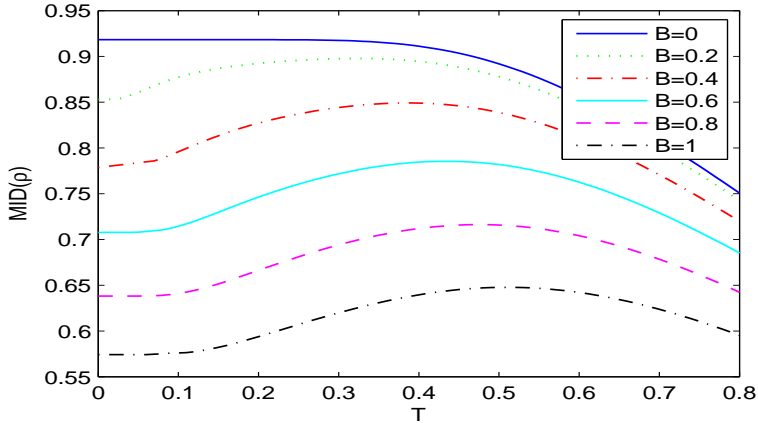}
\caption{(Color online) Quantum correlation measured by measurement-induced disturbance for any $B$ when $J=1$.}
\end{center}
\end{figure}

\section{Conclusions } Many people take it for granted that quantum correlation exists only in entangled state. By using measurement-induced disturbance we have investigated quantum correlation in a qutrit-qubit mixed spin chain. The dependences of measurement-induced disturbance on external magnetic field and spin-spin coupling are given in detail. More importantly, we have compared measurement-induced disturbance with quantum thermal entanglement measured by negativity and found no definite link between them. We find  the effect of temperature on measurement-induced disturbance is far weaker than on thermal negativity. Thermal negativity will experience a sudden transition when temperature approaches zero, but this will not happen for measurement-induced disturbance. There is no thermal concurrence for a ferromagnetic qutrit-qubit model, measurement-induced disturbance exists for both antiferromagnetic and ferromagnetic case. All results show that quantum entanglement is not same with quantum correlation. Quantum entanglement is only a special kind of quantum correlation and separable state can possess quantum correlation.

\section{Acknowledgements}
This work was supported by the National Science Foundation of China under Grants No. 10874013 and 10904165.

\end{document}